\documentclass[aps,prl,twocolumn,superscriptaddress]{revtex4-2}\usepackage[utf8]{inputenc}
\usepackage{amsmath}
\usepackage{bm}
\usepackage{amssymb,amsfonts,latexsym,fancyhdr,graphicx,epstopdf}
\usepackage{graphicx}
\usepackage{soul}
\setstcolor{red}
\usepackage{times}
\usepackage{mathptmx}
\usepackage[english]{babel}
\usepackage{graphicx}
\usepackage {subfigure}
\usepackage[pdftex,colorlinks=true,allcolors=blue]{hyperref}
\usepackage{xcolor}
\usepackage{braket}
\usepackage{verbatim}
\newcommand{\nico}[1]{\textcolor{red}{#1}}

\newcommand{\canc}[1]{}
\usepackage{feynmp}
\usepackage{amsmath,tkz-euclide}
\newcommand{\msc}[1]{\normalfont\textsc{#1}}

\begin{document}
\def\infnlnfcs{INFN, Gruppo collegato di Cosenza, I-87036 Rende (CS), Italy}
\def\fisunical{Dip. di Fisica, Universit\`{a} della Calabria,  I-87036 Rende (CS), Italy}

\title{Revealing Many-Body Phases at finite temperatures through Quantum Coherence Distribution}
%\title{Probability Distribution of Coherences as a Probe of Single and Many-Body Phases }
\author{A. Palamara}
\affiliation{\fisunical}
\affiliation{\infnlnfcs}
\email{antonio.palamara@unical.it}
\author{ N. Lo Gullo}
\affiliation{\fisunical}
\affiliation{\infnlnfcs}
\email{nicolino.logullo@unical.it}
\author{A. Sindona}
\affiliation{\fisunical}
\affiliation{\infnlnfcs}
\email{antonello.sindona@fis.unical.it}
\author{F. Plastina}
\affiliation{\fisunical}
\affiliation{\infnlnfcs}
\email{francesco.plastina@unical.it}

\begin{abstract}
We introduce a diagnostic framework based on the full probability distribution of quantum coherence to probe single-particle and many-body phases at finite temperature. We first apply this approach to the localization transition in the non-interacting Aubry–André model and then extend the analysis to its interacting counterpart, whose phase diagram is substantially richer. We show that the moments of the coherence distribution provide clear signatures of quantum phase transitions and capture the emergence of many-body phases. More importantly, we find that the entropy of the coherence distribution emerges as an even more powerful indicator, retaining clear signatures of the underlying quantum phases while remaining remarkably robust against increasing thermal fluctuations.

%We propose a novel diagnostic tool based on the probability distribution of quantum coherences to probe single-particle and many-body phases at finite temperature. Working within a two-point measurement framework, we first apply this approach to the localization transition in the non-interacting Aubry–André model and then extend it to its interacting version, where the phase diagram becomes substantially richer. Our analysis shows that the moments of the coherence probability distribution carry clear signatures of Quantum phase transitions and  phase diagrams, capturing the emergence of many-body phases. More importantly, we find that the entropy of the distribution not only retains these signatures, but is significantly more robust against increasing classical thermal fluctuations.

\end{abstract}

\maketitle

%\section{Introduction} 
\textbf{\textit{Introduction}} --     Quantum coherence, the defining feature of quantum superposition, underpins many of the distinctive phenomena of quantum mechanics. Over the past decades, the quantification \cite{PhysRevLett.113.140401,PhysRevA.91.042120,PhysRevA.92.042101} and manipulation of coherence have become central to the understanding of a broad range of physical processes, from quantum information protocols \cite{streltsov2017colloquium} to the thermodynamics of nonequilibrium quantum systems \cite{francica2019role,su2018heat,gour2022role,narasimhachar2015low,henao2018role,Rodrigues2024,santos2019role,qxmt-m3rk}, where coherence plays a key role in energy conversion, fluctuation relations, and irreversibility. More recently, quantum coherence has also emerged as a valuable tool for investigating many-body localized phases \cite{PhysRevA.110.022434,PhysRevB.100.224204,PhysRevB.111.054202,PhysRevB.102.045140}, excited-state quantum phase transitions \cite{wang2025characterizinglipkinmeshkovglickexcitedspectrum}, and topologically frustrated spin chains \cite{dl7m-8qyt}.

From the experimental perspective, ultracold atomic gases have become an exceptionally clean and coherently controllable platform for simulating condensed-matter phenomena  \cite{bloch2008many}. Optical lattice experiments, combined with the ability to tune interactions through Feshbach resonances \cite{lewenstein2007ultracold,courteille1998observation}, have enabled the exploration of a wide variety of quantum many-body phases and phase transitions, including those occurring in Bose–Hubbard systems, SU(N) Hamiltonians, and fermionic systems with random or quasiperiodic potentials \cite{doi:10.1126/science.aaa7432,CAPPONI201650,cazalilla2014ultracold,jaksch1998cold,Muniz2020,Roati2008,Greiner2002,Song2022}

While quantum phase transitions are, in principle, defined at zero temperature, finite-temperature effects are unavoidable in realistic experiments.
%\st{, since reaching absolute zero is prohibited by the third law of thermodynamics}.
As the temperature increases, thermal fluctuations progressively obscure the signatures of quantum criticality, replacing sharp critical lines with finite-width crossover regions \cite{PhysRevLett.109.160601, MatthiasVojta_2003,Palamara_2026,PhysRevA.110.062203,PhysRevB.89.165117,AAA1}. This makes the identification of quantum criticality at finite temperature a challenging task. To date, only a few quantities have been shown to retain sensitivity to quantum critical points in the presence of thermal fluctuations, including quantum discord \cite{PhysRevLett.105.095702,PhysRevA.83.062334,doi:10.1142/S021797921345032X} and quantum teleportation based schemes \cite{PhysRevA.111.012404,PhysRevA.107.052420}.

Motivated by these theoretical and experimental developments, we investigate the probability distribution of quantum coherence, a quantity rooted in nonequilibrium thermodynamics, as a diagnostic tool to probe quantum phases and phase transitions in finite-temperature systems, in both single-particle and many-body systems. Unlike previous coherence-based approaches, which rely primarily on average quantities, our framework exploits the full probability distribution, whose first moment recovers the relative entropy of coherence. This naturally provides access to higher-order moments, allowing us to characterize quantum fluctuations and their interplay with thermal fluctuations, which introduce classical uncertainty into the state. Beyond the information encoded in its moments, the entropy of the coherence distribution proves considerably more robust against increasing temperature than any individual moment, making it a reliable indicator of quantum criticality even in the presence of strong thermal fluctuations.

To demonstrate the effectiveness of our framework, we consider two paradigmatic models. We first investigate the non-interacting Aubry-André model \cite{ref7,ref8,Domínguez-Castro_2019}, where the localization transition provides an ideal benchmark for our approach. We then extend the analysis to the many-body Aubry-André model with nearest-neighbor interactions \cite{naldesi2016detecting}, which exhibits a much richer phase diagram. In both cases, we show that the coherence distribution captures the underlying ground-state properties even at finite temperature, with its entropy providing the most robust indicator against thermal effects.

%The remainder of the paper is organized as follows. Section \ref{secPC} introduces the theoretical framework and the probability distribution of quantum coherence. Sections \ref{secAA} and \ref{secAAi} present its application to the non-interacting and interacting Aubry-André models, respectively. Finally, Sec. \ref{secconclu} summarizes our conclusions and discusses possible future directions.

%\section{Probability distribution of coherences}
\textbf{\textit{{Probability distribution of coherences}}} --
\label{secPC}
%Let us consider a generic Hamiltonian $\hat{{H}}$, that exhibits a quantum phase transition and has eigenvectors \(\ket{\msc{E}_n}\).
Inspired by the stochastic approach to quantum thermodynamics, we introduce the probability distribution of coherence. In the following, we show how this quantity can be used to detect signatures of both single-particle and many-body localized phases, even at finite temperature, where the physical state is generally mixed. 

Consider a basis $\{\ket{\alpha}\}$, spanning the Hilbert space of the system (possibly many-body). The coherence of a generic state $\hat{\rho}$ is quantified by the relative entropy of coherence \cite{PhysRevLett.113.140401}
\begin{equation}
    {C}[\hat{\rho}]= {S}(\hat{  \rho}||\Delta_{\alpha}[ \hat{\rho}])={S}(\Delta_{\alpha}[ \hat{\rho}])-{S}(\hat{\rho})
\end{equation}

Here, $\Delta_{\alpha}[ \cdot]$ represents the dephasing operator in the $\{ \ket{\alpha} \}$ basis:

\begin{equation}
    \Delta_{\alpha}[\hat{\rho}]= \sum_{\alpha}\braket{\alpha| \hat{\rho} | \alpha} \ket{\alpha}\bra{\alpha}=\sum_{\alpha}p_\alpha \ket{\alpha}\bra{\alpha},
\end{equation}
where $p_{\alpha}=\braket{\alpha| \hat{\rho} | \alpha}$ is a diagonal element of $\hat{\rho}$ in the chosen basis.
Thus, coherence, evaluated as the relative entropy between the state $\hat{\rho}$ and $\Delta_{\alpha}[ \hat{\rho}]$, can be interpreted as the entropy increase induced by the dephasing operation.

In the context of quantum thermodynamics, coherence generated during a nonequilibrium process can be associated with a stochastic variable ~\cite{francica2019role}. Building on this idea, we define a probability distribution of coherence whose moments provide access to physically meaningful quantities displaying signatures of quantum criticality.
The emergence of critical signatures relies on an appropriate choice of the basis in which coherence is evaluated; this will be made explicit below.

Let us consider a generic state $\hat{\rho}$, whose spectral decomposition reads
\begin{equation}
\hat{\rho}= \sum_{n} p_n \ket{r_n}\bra{r_n}.
\end{equation}
The probability distribution of coherence is constructed using a two-point measurement (TPM) protocol. The first measurement is performed in the basis of the eigenstates of the state $\hat{\rho}$. For an equilibrium thermal state, this simply corresponds to an energy measurement. A second projective measurement is then performed in the basis $\{\ket{\alpha}\}$.

The stochastic coherence variable $c$, associated with the measurement outcomes is defined as 
\begin{equation}
    c_{n\alpha}= \ln p_n- \ln p_\alpha .
    \label{eq4}
\end{equation}
Collecting all possible measurement outcomes yields the probability distribution
\begin{equation}
    P(c)= \sum_{n,\alpha} p_n|\braket{r_n|\alpha}|^2 \delta(c-c_{n\alpha}).
    \label{eq3}
\end{equation}
The statistical fluctuations arising from the measurement outcomes, together with those originating from the non-commutativity of the two measurements, are encoded in the joint probability $p_n|\braket{{r}_n|\alpha}|^2$,  constrained by the fluctuation relation:
\begin{equation}
    \braket{e^{-c}}=1.
\end{equation}
A key property of the coherence distribution is that its first moment exactly reproduces the relative entropy of coherence; namely: $\langle c \rangle = \int c \, P(c) dc \equiv C[\hat{\rho}]$.

Higher-order moments of the distribution can be obtained from the characteristic function $\chi_c(\nu)$. Specifically, the $n$-th moment follows from $\braket{c^n} = (-i)^n\partial^n_\nu \chi_c(\nu)|_{\nu=0}$, where  the characteristic function is
\begin{align}
    \chi_c (\nu)= &\int e^{i\nu c}P(c)dc = \sum_{n,\alpha} \frac{p_n^{i\nu+1}}{p_\alpha^{i\nu}}|\braket{{r}_n|\alpha}|^2=\\
    =&\mathrm{Tr}\big{\{}\hat{\rho}e^{i\nu\ln\hat{\rho}}e^{-i\nu\ln \Delta_{   \alpha}[\hat{\rho}]}\big{\}}.
\end{align} 

Motivated by the analogous construction for the probability distribution of work \cite{PhysRevResearch.5.L022010}, we also introduce the entropy of the coherence distribution. While the moments characterize specific statistical properties of the distribution, its entropy provides a global measure of the information encoded in the stochastic coherence. It reads:

\begin{equation}
\msc{H}_c=-\sum_{c} P(c)\ln P(c)=
-\sum_{n,\alpha} p_n |\braket{r_n|\alpha}|^2\ln{P(c_{n,\alpha})}
\end{equation}
Here, $P(c_{n,\alpha})$ denotes the probability that the stochastic coherence takes the value $c_{n,\alpha}$, namely $P(c_{n,\alpha})=\sum_{(m,\beta) \in I_{n,\alpha}} p_m \, |\langle r_m|\beta\rangle|^2$, with the set ${I}_{n,\alpha} $ contains all pairs $(m,\beta)$ yielding the same value of the stochastic coherence $c_{n,\alpha}$.

By summing and subtracting the term $\sum_{n,\alpha} p_n |\braket{r_n|\alpha}|^2\ln{p_n |\braket{r_n|\alpha}|^2}$ we can rewrite: 
\begin{align}
\msc{H}_{c} &= {S}(\hat{\rho})+\sum_{n}p_n {C}[\ket{r_n}\bra{r_n}]\\
&-\Delta(\{p_n |\braket{r_n|\alpha}|^2\},\{P(c_{n,\alpha})\}).\notag
\label{hh2}
\end{align}

Thus, the entropy of the coherence distribution is the sum of three contributions: the von Neumann entropy of the mixed state, the average coherence of its eigenstates, and a correction accounting for the degeneracy of the stochastic coherence through the term $\Delta =\sum_{n,\alpha} p_n|\braket{r_n|\alpha}|^2\ln[\frac{P(c_{n,\alpha})}{p_n|\braket{r_n|\alpha}|^2}]$. The above expression highlights the interplay between thermal mixedness and quantum coherence thus providing the basis for the finite-temperature analysis developed in the following sections.
Moreover, since $P(c_{n,\alpha}) \geq p_n|\braket{r_n|\alpha}|^2$, it follows that $\Delta\geq 0$. Therefore, we have
 that $\msc{H}_{c}\leq\msc{H}^u_{c}$ where we defined the upper bound:
\begin{equation}
\msc{H}^u_{c} = {S}(\hat{\rho})+\sum_{n}p_n {C}[\ket{r_n}\bra{r_n}].
\label{hh}
\end{equation}
%\here

We now apply this framework to two paradigmatic localization models. We first consider the non-interacting Aubry-André model and subsequently its interacting extension, showing that the coherence distribution (and particularly its entropy) retains clear signatures of the underlying quantum phases even at finite temperatures.

\begin{figure*}[htbp]
  \centering
  \includegraphics[width=0.9\textwidth]{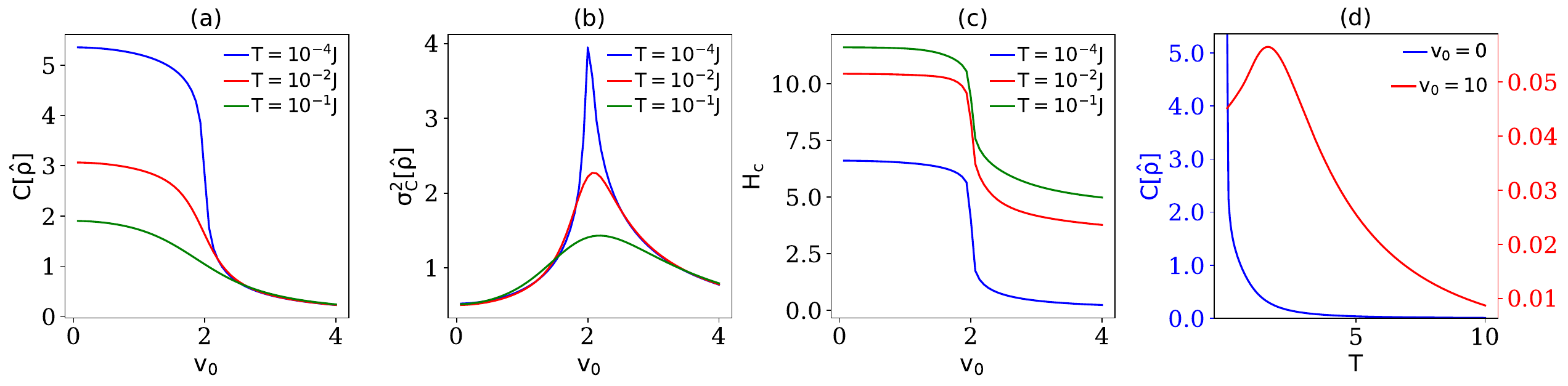}
\caption{Finite-temperature behavior of the coherence probability distribution in the non-interacting Aubry-André model. (a) Coherence $C[\hat{\rho}_\beta]$ and (b) variance of the coherence probability distribution, $\sigma^2_C[\hat{\rho}_\beta]$, as functions of the quasiperiodic potential strength, for three different temperatures.
(c) Entropy of the coherence probability distribution, $\msc{H}_c$, as a function of $\msc{v}_0$. (d) Temperature dependence of the coherence $C[\hat{\rho}_\beta]$  deep in the delocalized and localized phases. All results are obtained for a lattice of
$\msc{N}=10^{3}$ sites.}
  \label{fig1}
\end{figure*}
%\section{Localization in the Aubry-André (AA) model}\label{secAA}
\textbf{\textit{Localization in the Aubry-André (AA) model}} --
We begin with the non-interacting Aubry-André model, \cite{ref7,ref8,Domínguez-Castro_2019}, a paradigmatic system exhibiting a localization transition. It describes a single particle on a one-dimensional lattice subjected to a quasiperiodic potential. The Hamiltonian reads :

\begin{equation}
   \hat{H}_{AA}=-J\sum^{\msc{N}-1}_{i=1}( \hat{c}^\dagger_i\hat{c}_{i+1}+h.c)+ J \msc{v}_0\sum^{\msc{N}}_{i=1} \cos{(2\pi\gamma i )} \hat{c}^\dagger_i\hat{c}_{i},
\end{equation}
where we assumed open boundary conditions. Here, $\hat{c}^\dagger(\hat{c})$ are the fermion creation (annihilation) operators of a particle at the $i$-th site; $J$ is the hopping amplitude that we take, from now on, as our energy (and temperature) unit. $\msc{v}_0$ is the strength of the quasiperiodic potential, and $\gamma = \frac{\sqrt{5}-1}{2}$ is an incommensurate modulation parameter. The AA model features a single-particle localization-delocalization transition due to the presence of the external quasiperiodic potential, which mimics a correlated disorder. Unlike single-particle systems subjected to uncorrelated disorder, the AA model displays a transition from a delocalized to a localized phase at a finite value of the external potential amplitude. Specifically, for $\msc{v}_0 < 2$, all of the energy eigenvectors are delocalized across the whole chain; while, for $\msc{v}_0 > 2$, they become exponentially localized.

To investigate the localization transition at finite temperature using the coherence probability distribution introduced in the previous section, 
we consider the thermal equilibrium state with inverse temperature $\beta = 1/T$ (throughout this work, $T$ includes the Boltzmann constant and is expressed in units of the hopping amplitude J $J$)
 \begin{equation}
     \hat{\rho}_\beta = \frac{e^{-\beta \hat{H}_{AA}}}{\msc{Z}} \hspace{0.5cm} \msc{Z}= Tr\{e^{-\beta \hat{H}_{AA}}\}.
 \end{equation}

As discussed previously, the emergence of critical signatures depends on the choice of the basis in which coherence is evaluated.
Since the localization transition of the Aubry–André model is naturally characterized in real space, we evaluate coherence in the local basis $\{\ket{i}\}$, where $i = 1,  \ldots ,\msc{N}$ labels the lattice sites.
%\canc{We choose to evaluate the coherences, i.e., perform the second measurement to construct the probability distribution, in the basis $\{\ket{i}\}$, where $i$ is the index labeling the $i$-th site.}
\begin{figure}[htbp]
  \centering
     \includegraphics[width=0.25\textwidth]{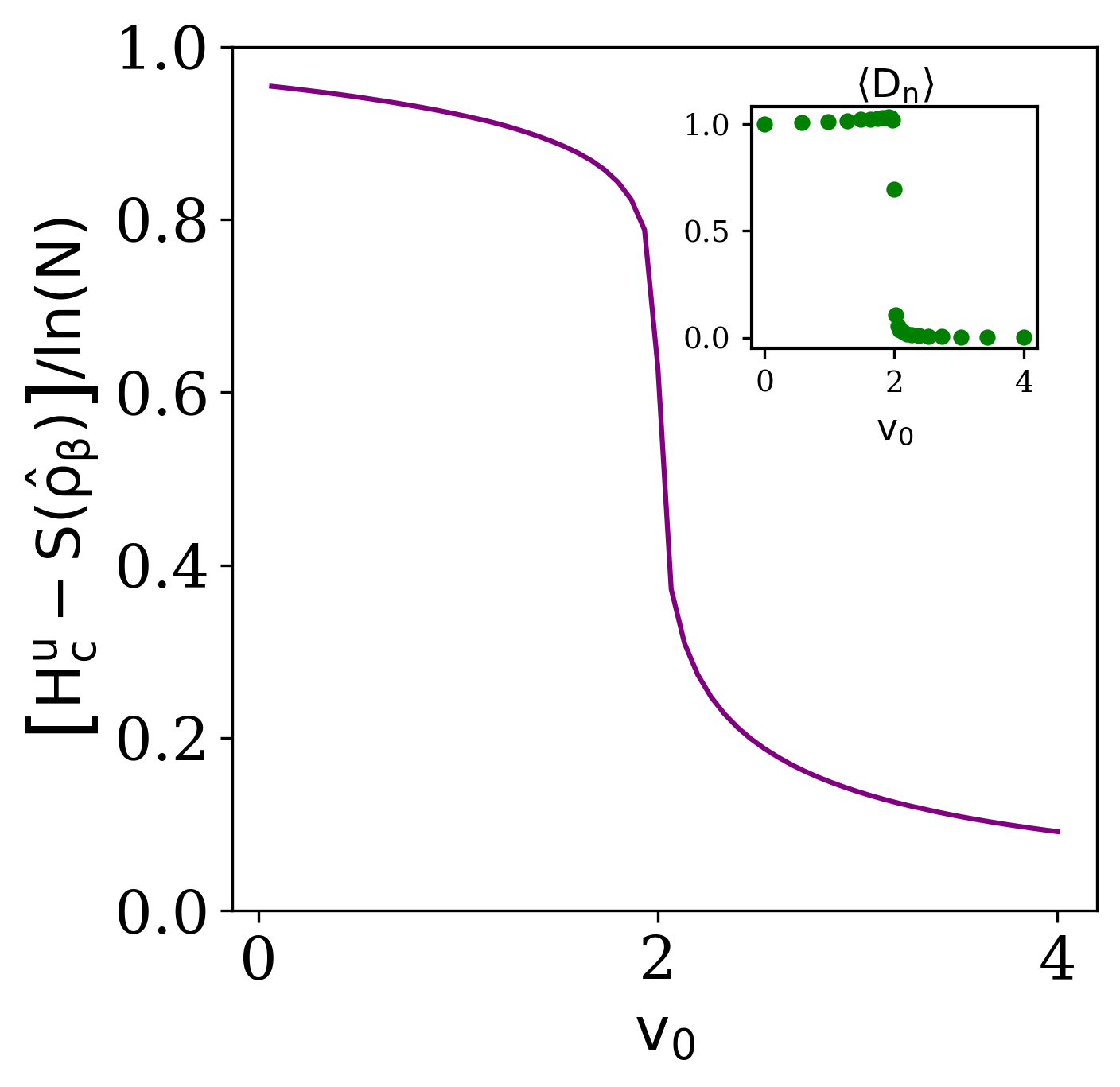}
\caption{Contribution of the individual eigenstate coherences to the uncollected entropy $\msc{H}^u_{c}$ in the high-temperature limit, as a function of the quasiperiodic potential strength $\msc{v}_0$, for a lattice of $\msc{N}=10^3$ sites. The inset shows the average fractal dimension $\braket{D_n}$, obtained from finite-size scaling of the Aubry-André model.}
  \label{fig2}
\end{figure}
In this basis, all the moments of the coherence distribution exhibit clear signatures of the localization transition, which remain observable up to finite temperatures. The first moment of the distribution, corresponding to the average coherence, is larger in the extended phase and decreases with increasing temperature, while it vanishes in the fully localized limit, $\msc{v}_0 \to \infty$, where $[\hat{H},\ket{i}\bra{i}] = 0$ for every lattice site.

As shown in Fig. \ref{fig1} (a), increasing the temperature reduces the coherence in the delocalized phase and progressively broadens the crossover region around the critical point $\msc{v}_0 =2$.

Figure \ref{fig1} (b) shows the second central moment of the coherence distribution, namely the variance $\sigma^2_{{C}}[\hat{\rho}] = \braket{c^2} - \braket{c}^2$. A pronounced peak develops at the localization transition, reflecting the critical nature of the eigenstates of the Hamiltonian. The variance receives contributions from both thermal fluctuations, encoded in the populations of the energy eigenstates, and quantum fluctuations arising from the non-commutativity between the Hamiltonian and the local projectors.
As the temperature increases (i.e., as $T$ approaches $J$), the thermal contribution progressively dominates the variance, causing the critical peak to broaden and eventually suppressing the signatures of the localization transition in the coherence moments.

Figure \ref{fig1} (d) shows the temperature dependence of the average coherence in both the delocalized and localized phases.  In the delocalized regime, the temperature dependence is almost entirely governed by the von Neumann entropy of the thermal state. 
By contrast, coherence is strongly suppressed in the localized phase, as already evident from Fig.  \ref{fig1} (a). In this regime, its temperature dependence results from the competition between the von Neumann entropy of the thermal state and that of its dephased counterpart in the local basis. This interplay gives rise to a shallow maximum at intermediate temperatures, whose amplitude decreases progressively with increasing $\msc{v}_0$.

Overall, the first two moments of the coherence distribution retain clear signatures of the localization transition only at sufficiently low temperatures. As thermal fluctuations increase, these signatures become progressively washed out. 

A markedly different picture emerges  when considering a global property of the coherence distribution, namely its entropy, Fig. \ref{fig1} (c). Unlike the individual moments, the entropy retains a clear signature of the localization transition even at temperatures for which their critical behavior has essentially disappeared. This robustness reflects the fact that the entropy probes the coherence distribution as a whole, rather than a single statistical moment. As a consequence, it remains sensitive to the changes in the distribution induced by the localization transition even when the critical signatures in the individual moments are strongly suppressed by thermal fluctuations. These results show that the entropy of the coherence distribution provides a significantly more robust indicator of the underlying quantum phase than any individual moment.

As shown in Fig. \ref{fig1} (c), the entropy is larger in the delocalized phase, where the coherence distribution is broader, and decreases sharply upon entering the localized regime, where the distribution becomes progressively narrower. Remarkably, the sharp change across the localization transition remains clearly visible over the entire temperature range considered, demonstrating that the entropy of the coherence distribution provides a significantly more robust indicator of the underlying quantum phase than any individual moment.

To gain further insight into the origin of this behavior, we now consider the upper bound of the entropy of the  coherence distribution,
\begin{equation}
\msc{H}^u_{c} = {S}(\hat{\rho}_\beta)+\sum_{n}\frac{e^{-\beta \msc{E}_n }} {\msc{Z}}C(\ket{\msc{E}_n}\bra{\msc{E}_n}). \label{HHu}
\end{equation}
where ${\ket{\msc{E}_n}}$ denotes the eigenstate of $\hat{H}$ with eigenvalue ${\msc{E}_n}$. 

The second term of Eq. (\ref{HHu}) is the thermal average of the coherence of the individual energy eigenstates. Since these are pure states, their coherence reduces to the Shannon entropy of the corresponding probability distribution in the local basis, $C(\ket{\msc{E}_n}\bra{\msc{E}_n})=-\sum_i |\braket{i|\msc{E}_n}|^2\ln{|\braket{i|\msc{E}_n}|^2}$.
It is interesting to notice that the asymptotic scaling of the coherence with the system size provides direct information on the spatial structure of the eigenstates. Indeed, for fully delocalized states, $C(\ket{\msc{E}_n}\bra{\msc{E}_n})\simeq\ln{\msc{N}}$, whereas it vanishes for perfectly localized states. Close to the localization transition, where the eigenstates are multifractal, the coherence scales as $C(\ket{\msc{E}_n}\bra{\msc{E}_n})\simeq D_n\ln{\msc{N}}$, with $D_n$ the fractal dimension of the n-th eigenstate \cite{PhysRevLett.123.180601}.
Consequently, the contribution arising from the coherences of the individual eigenstates to the uncollected entropy exhibits the following asymptotic scaling with the system size: $\msc{H}^u_{c}-{S}(\hat{\rho}_\beta)= \braket{D_n}\ln(\msc{N})$, where $\braket{D_n}=\sum_n \frac{e^{-\beta \msc{E}_n }}{\msc{Z}} D_n$ is the thermal average of the fractal dimension of the eigenstates. Therefore, in the high-temperature limit ($\beta \rightarrow 0$), $\braket{D_n}$ approaches the average fractal dimension of the spectrum. Figure \ref{fig2} confirms this prediction by comparing the contribution of the eigenstate coherences to the uncollected entropy with the average fractal dimension extracted from finite-size scaling. The excellent agreement demonstrates that the entropy of the coherence distribution directly captures the fractal properties of the eigenstates across the localization transition.

%\section{Interacting Aubry-André model}\label{secAAi}
\textbf{\textit{Interacting Aubry-André model}} -- Having shown that the coherence distribution reveals the single-particle localization transition, we now investigate whether the same framework can identify many-body phases in the interacting Aubry-André model.

We consider $\msc{N}_p$ spinless interacting fermions subject to the Aubry-André potential, with nearest-neighbor density-density interactions. The Hamiltonian reads
 
\begin{equation}
\hat{H}= \hat{H}_{AA}+ U\sum^{\msc{N}-1}_{i=1}\hat{n}_i\hat{n}_{i+1}.
\label{HH}
\end{equation}
Here, $\hat{n}_i$ is the local number operator, while $U$ is the interaction strength. We consider half filling ($\msc{N}_p= \frac{\msc{N}}{2}$) and, as in the non-interacting case,  open boundary conditions. Owing to the exponential growth of the Hilbert space, the numerical results shown below are obtained for $\msc{N}=12$  (to maintain a manageable computational complexity). 
\begin{figure}[htbp]
  \centering
     \includegraphics[width=0.5\textwidth]{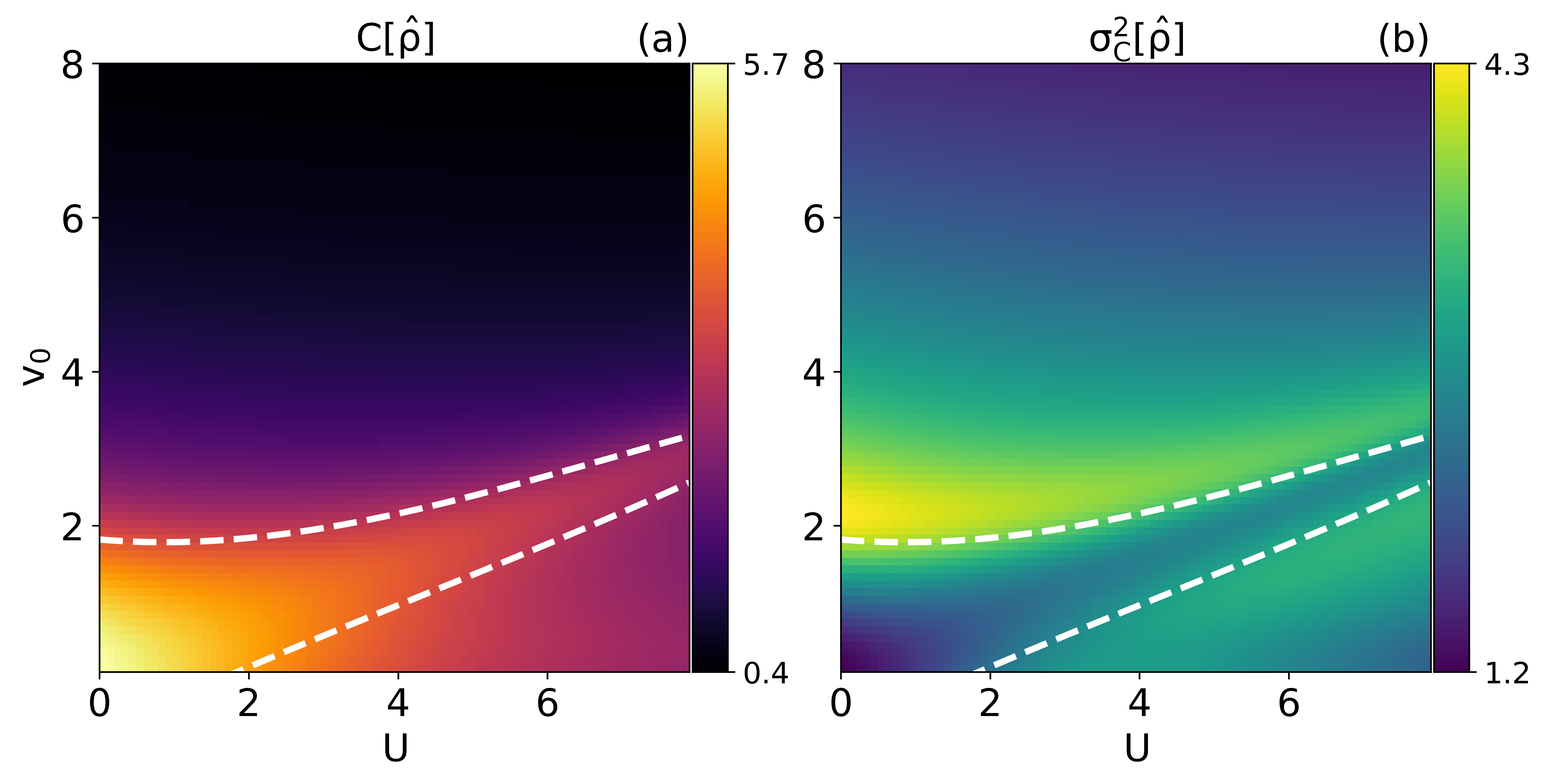}
  \caption{Ground-state coherence probability distribution in the interacting Aubry-André model. (a) Coherence $C[\hat{\rho}]$ and (b) variance $\sigma^2_c[\hat{\rho}]$, as functions of the interaction strength $U$ and quasiperiodic potential strength $\msc{v}_0$.
 %Panel (c): Coherence of the single-particle density matrix $C[\hat{\rho}_{ij}]$.
The dashed white lines indicate the phase boundaries obtained from the maxima of $\frac{\partial C[\hat{\rho}]}{\partial \msc{v}_0}$ and $\frac{\partial C[\hat{\rho}]}{\partial U}$, respectively, and are included as guides to the eye separating the MP, ABI, and CDW phases.}
  \label{fig3}
\end{figure}

%\here
\canc{The interplay between the quasiperiodic potential and interactions is known to give rise to a many-body localized (MBL) phase, as demonstrated both theoretically \cite{PhysRevLett.115.180401,PhysRevA.95.033605, PhysRevB.87.134202,naldesi2016detecting, bera2017one, lev2017transport} and experimentally \cite{PhysRevLett.119.260401,schreiber2015observation}. Moreover, interactions enrich the phase diagram beyond the MBL regime, leading to the emergence of a metallic phase (MP) and a charge-density-wave (CDW) phase.

For quasiperiodic systems, it has been that the interplay between the incommensurate external potential and the interactions gives rise to a many-body localized (MBL) phase. Moreover, in Ref. \cite{PhysRevB.101.144303} it has been shown that the slowing down of the dynamics observed by L\"uschen et al., in Ref.  \cite{PhysRevLett.119.260401} is due to the persistence of the critical nature of the single particle-spectrum in the presence of interactions. Beyond the MBL phase, the competition between particle-particle interaction and the quasiperiodic potential significantly enriches the phase diagram, leading to the emergence of a metallic phase (MP) and a charge density wave (CDW) phase \cite{naldesi2016detecting}. }

The interplay between the quasiperiodic potential and interactions is known to give rise to interesting effects such as a many-body localized (MBL) phase, as demonstrated both theoretically \cite{PhysRevLett.115.180401,PhysRevA.95.033605, PhysRevB.87.134202,naldesi2016detecting, bera2017one, lev2017transport} and experimentally \cite{PhysRevLett.119.260401,doi:10.1126/science.aaa7432}, the emergence of a metallic phase (MP) and a charge-density-wave (CDW) phase  \cite{naldesi2016detecting}, the slowing down of the dynamics \cite{PhysRevLett.119.260401} which has been shown~\cite{PhysRevB.101.144303} to arise from the critical nature of the single-particle spectrum.

%\here

Our central question is whether the coherence probability distribution can distinguish these many-body phases and retain clear signatures of their boundaries.

As in the non-interacting case, we begin by considering the thermal equilibrium state
\begin{equation}
     \hat{\rho}_\beta = \frac{e^{-\beta \hat{H}}}{\msc{Z}} \hspace{0.5cm} \msc{Z}= Tr\{e^{-\beta \hat{H}}\}.
     \label{thermalI}
 \end{equation}
We first analyze the ground-state limit $\beta \xrightarrow{} \infty$ by studying the first two moments of the coherence distribution. 
%As discussed in Sec. \ref{secPC}, the emergence of critical signatures depends on the choice of the reference basis. 
Since the phases of the interacting model are naturally characterized in terms of many-body occupation patterns, we evaluate coherence in the basis of many-body real-space configurations $\{\ket{\{n_i\}_J}\}$ where $J = 1,  \ldots ,\msc{D}$. Here $\msc{D}= \binom{\msc{N}}{\msc{N}_p}$ is the Hilbert space dimension.

\begin{figure}[htbp]
  \centering
     \includegraphics[width=0.5\textwidth]{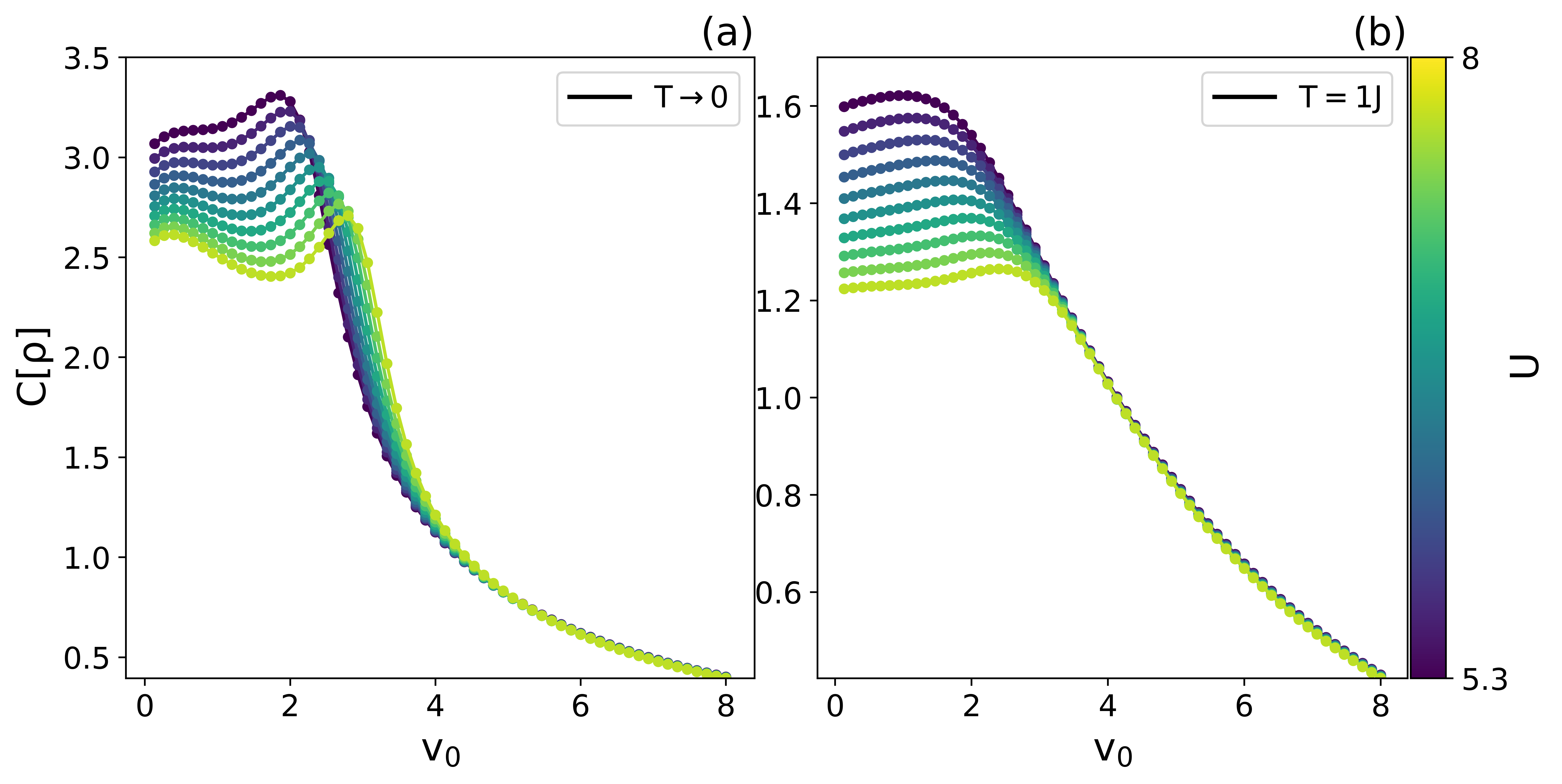}

\caption{Effect of finite temperature on the coherence in the interacting Aubry-André model. (a) Ground-state coherence as a function of the quasiperiodic potential strength $\msc{v}_0$ for different interaction strenght $U$. (b) Thermal-state coherence at temperature $T=1J$  as a function of $\msc{v}_0$ for the same values of $U$. }
  \label{fig4}
\end{figure}
The expected behavior of the coherence moments follows from the nature of the three phases. In the metallic phase (MP), the ground state is delocalized over the many-body configuration basis, and the coherence is therefore expected to approach its maximum value with a coherence distribution sharply peaked around this value, resulting in a small variance.

At the opposite extreme, the coherence is strongly suppressed in the Anderson band insulator (ABI), where $ \msc{v}_0/U \gg 1$ , while it assumes an intermediate, nearly constant value in the charge-density-wave (CDW) phase, corresponding to
$ \msc{v}_0/U \ll 1$. Consequently, the variance is expected to be largest in the crossover regions separating these phases, where the coherence distribution broadens due to the competition between different many-body configurations.

This behavior closely parallels the non-interacting Aubry-André model, with the many-body configuration basis replacing the local basis. Figure \ref{fig3} (a) and (b) confirm this physical picture. The coherence exhibits the expected behavior across the three phases, while the variance develops a pronounced peak as the system approaches the ABI phase, reflecting the persistence of the delocalized-localized transition even in the presence of moderate interactions \cite{PhysRevLett.115.180401}. 

Overall, the behavior of the first two moments demonstrates that, with an appropriate choice of basis, the coherence probability distribution provides a useful characterization of the many-body phases at zero temperature.

\begin{figure}[htbp]
  \centering
     \includegraphics[width=0.5\textwidth]{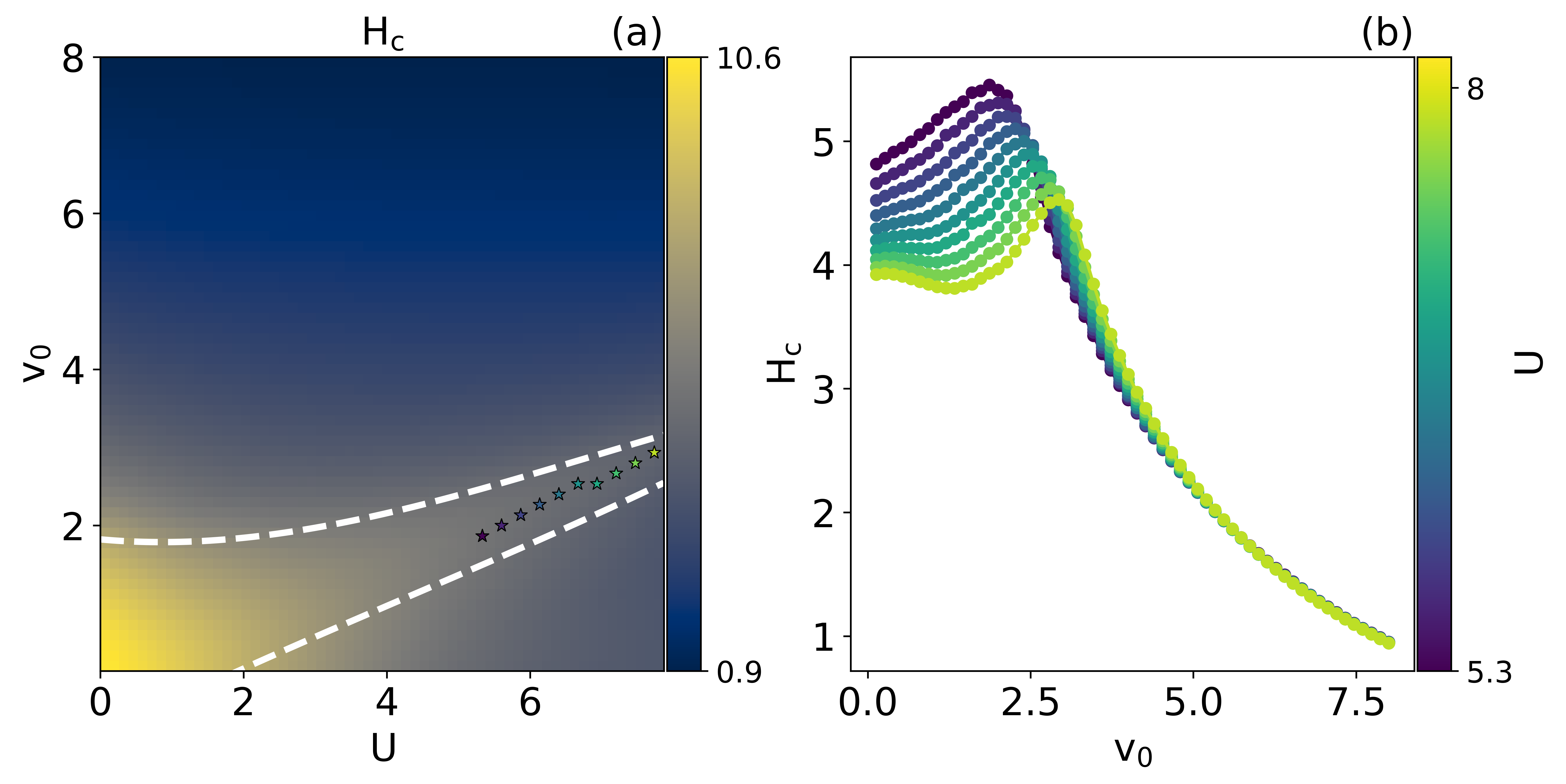}
\caption{Entropy of the coherence probability distribution in the interacting Aubry-André model at finite temperature. (a) Entropy $\msc{H}_c$ as a function of the quasiperiodic potential strength $\msc{v}_0$ and interaction strength $U$ for a thermal state with $\beta=1J$. The dashed white lines indicate the phase boundaries shown in Fig. \ref{fig3} and are included as guides to the eye.  Panel (b):  Entropy as a function of $\msc{v}_0$ for different interaction strengths $U$. 
  }
  \label{fig5}
\end{figure}

We now turn to the effects of finite temperature.

Figure \ref{fig4} compares the coherence of the ground state with that of the thermal equilibrium state at inverse temperature $\beta J = 1$, for different interaction strengths as a function of $\msc{v}_0$.

As in the non-interacting model, thermal fluctuations progressively wash out the signatures of the many-body phases. In particular, the coherence peak associated with the metallic phase becomes substantially broader, making the phase boundaries increasingly difficult to identify from the first moment alone.

These observations naturally motivate the analysis of the entropy of the coherence distribution, as in the single-particle case. 

As shown in Fig. ~\ref{fig5} (a), unlike the individual moments, the entropy continues to provide a clear identification of the three phases of the interacting Aubry-André model throughout the $(U,\msc{v}_0)$ phase diagram, even at finite temperature. This robustness indicates that the entropy is significantly less sensitive to thermal fluctuations than the individual moments of the coherence distribution.

Remarkably, the entropy of the probability distribution continues to provide a clear identification of the three phases of the interacting Aubry--André model in the  plane (see the first panel of Fig.~\ref{fig5}) even at finite temperature. This suggests that its dependence on temperature, or equivalently on the von Neumann entropy of the thermal state, is significantly weaker than that exhibited by the individual moments of the probability distribution.

%\section{Conclusion and outlook}
\label{secconclu}
\textbf{\textit{Conclusion and outlook}} --
In this work, we introduced and investigated the probability distribution of quantum coherence as a tool to characterize quantum phases at finite temperature. Our main result is that, while the individual moments of the distribution become progressively less informative as thermal fluctuations increase, its entropy remains a robust indicator of the underlying quantum phases.

We demonstrated this framework in both the non-interacting and interacting Aubry-André models, showing that the average coherence reaches a maximum in the delocalized phase and a minimum in the strongly localized regime, while the variance exhibits a peak near the critical point. Additional signatures are revealed in the presence of interactions, where a more complex phase diagram is displayed, comprising ABI, MP and CDW phases.

In both the single and many-body models, the coherence moments capture the localization properties at low temperature but gradually lose their critical signatures as thermal fluctuations increase. By contrast, the entropy of the coherence distribution retains clear signatures of the different phases across a much broader temperature range.

More broadly, our results suggest that the coherence distribution provides a natural framework for disentangling quantum and thermal fluctuations in finite-temperature many-body systems.

These results highlight the importance of considering the full coherence distribution rather than individual statistical moments when investigating finite-temperature quantum systems.

Several directions naturally follow from this work. An important extension concerns nonequilibrium dynamics, where the time evolution of the coherence distribution may provide new insights into dynamical phase transitions. Equally important is the development of experimentally accessible protocols for reconstructing the coherence distribution, for example through quench-based schemes, which would open the way to direct experimental tests of our results.

%\section{Acknowledgements}
\textbf{\textit{Acknowledgements}} --This research was supported by \emph{Centro Nazionale di Ricerca in High-Performance Computing, Big Data and Quantum Computing}, PNRR 4 2 1.4, CI CN00000013, CUP H23C22000360005, and by the PNRR MUR project PE0000023-NQSTI through the secondary projects “QuCADD” and “ThAnQ”.

\bibliography{bib}

@article{PhysRevLett.113.140401,
  author  = {Baumgratz, T. and Cramer, M. and Plenio, M. B.},
  title   = {Quantifying Coherence},
  journal = {Phys. Rev. Lett.},
  volume  = {113},
  pages   = {140401},
  year    = {2014},
  doi     = {10.1103/PhysRevLett.113.140401}
}

@article{PhysRevA.91.042120,
  author  = {Shao, L.-H. and Xi, Z. and Fan, H. and Li, Y.},
  title   = {Fidelity and trace-norm distances for quantifying coherence},
  journal = {Phys. Rev. A},
  volume  = {91},
  pages   = {042120},
  year    = {2015},
  doi     = {10.1103/PhysRevA.91.042120}
}

@article{PhysRevA.92.042101,
  author  = {Cheng, S. and Hall, M. J. W.},
  title   = {Complementarity relations for quantum coherence},
  journal = {Phys. Rev. A},
  volume  = {92},
  pages   = {042101},
  year    = {2015},
  doi     = {10.1103/PhysRevA.92.042101}
}

@article{streltsov2017colloquium,
  author  = {Streltsov, A. and Adesso, G. and Plenio, M. B.},
  title   = {Colloquium: Quantum coherence as a resource},
  journal = {Rev. Mod. Phys.},
  volume  = {89},
  pages   = {041003},
  year    = {2017},
  doi     = {10.1103/RevModPhys.89.041003}
}

@article{francica2019role,
  author  = {Francica, G. and Goold, J. and Plastina, F.},
  title   = {Role of coherence in the nonequilibrium thermodynamics of quantum systems},
  journal = {Phys. Rev. E},
  volume  = {99},
  pages   = {042105},
  year    = {2019},
  doi     = {10.1103/PhysRevE.99.042105}
}

@article{su2018heat,
  author  = {Su, S. and Chen, J. and Ma, Y. and Chen, J. and Sun, C.},
  title   = {The heat and work of quantum thermodynamic processes with quantum coherence},
  journal = {Chin. Phys. B},
  volume  = {27},
  pages   = {060502},
  year    = {2018},
  doi     = {10.1088/1674-1056/27/6/060502}
}

@article{gour2022role,
  author  = {Gour, G.},
  title   = {Role of quantum coherence in thermodynamics},
  journal = {PRX Quantum},
  volume  = {3},
  pages   = {040323},
  year    = {2022},
  doi     = {10.1103/PRXQuantum.3.040323}
}

@article{narasimhachar2015low,
  author  = {Narasimhachar, V. and Gour, G.},
  title   = {Low-temperature thermodynamics with quantum coherence},
  journal = {Nat. Commun.},
  volume  = {6},
  pages   = {7689},
  year    = {2015},
  doi     = {10.1038/ncomms8689}
}

@article{henao2018role,
  author  = {Henao, I. and Serra, R. M.},
  title   = {Role of quantum coherence in the thermodynamics of energy transfer},
  journal = {Phys. Rev. E},
  volume  = {97},
  pages   = {062105},
  year    = {2018},
  doi     = {10.1103/PhysRevE.97.062105}
}

@article{santos2019role,
  author  = {Santos, J. P. and C\'{e}leri, L. C. and Landi, G. T. and Paternostro, M.},
  title   = {The role of quantum coherence in non-equilibrium entropy production},
  journal = {npj Quantum Inf.},
  volume  = {5},
  pages   = {23},
  year    = {2019},
  doi     = {10.1038/s41534-019-0138-y}
}

@article{PhysRevA.110.022434,
  author  = {Chen, J.-J. and Xu, K. and Ren, L.-H. and Zhang, Y.-R. and Fan, H.},
  title   = {Dynamics of quantum coherence in many-body localized systems},
  journal = {Phys. Rev. A},
  volume  = {110},
  pages   = {022434},
  year    = {2024},
  doi     = {10.1103/PhysRevA.110.022434}
}

@article{PhysRevB.100.224204,
  author  = {Styliaris, G. and Anand, N. and Campos Venuti, L. and Zanardi, P.},
  title   = {Quantum coherence and the localization transition},
  journal = {Phys. Rev. B},
  volume  = {100},
  pages   = {224204},
  year    = {2019},
  doi     = {10.1103/PhysRevB.100.224204}
}

@article{PhysRevB.111.054202,
  author  = {Garg, A. and Pati, A. K.},
  title   = {Trade-off relations between quantum coherence and measure of many-body localization},
  journal = {Phys. Rev. B},
  volume  = {111},
  pages   = {054202},
  year    = {2025},
  doi     = {10.1103/PhysRevB.111.054202}
}

@article{PhysRevB.102.045140,
  author  = {Dhara, S. and Hamma, A. and Mucciolo, E. R.},
  title   = {Quantum coherence in ergodic and many-body localized systems},
  journal = {Phys. Rev. B},
  volume  = {102},
  pages   = {045140},
  year    = {2020},
  doi     = {10.1103/PhysRevB.102.045140}
}

@misc{wang2025characterizinglipkinmeshkovglickexcitedspectrum,
  author        = {Wang, Q. and P\'{e}rez-Bernal, F.},
  title         = {Characterizing the Lipkin-Meshkov-Glick excited spectrum through the quantum coherence spectrum},
  year          = {2025},
  eprint        = {2209.02510},
  archivePrefix = {arXiv},
  primaryClass  = {quant-ph}
}

@article{bloch2008many,
  author  = {Bloch, I. and Dalibard, J. and Zwerger, W.},
  title   = {Many-body physics with ultracold gases},
  journal = {Rev. Mod. Phys.},
  volume  = {80},
  pages   = {885--964},
  year    = {2008},
  doi     = {10.1103/RevModPhys.80.885}
}

@article{lewenstein2007ultracold,
  author  = {Lewenstein, M. and Sanpera, A. and Ahufinger, V. and Damski, B. and Sen(De), A. and Sen, U.},
  title   = {Ultracold atomic gases in optical lattices: mimicking condensed matter physics and beyond},
  journal = {Adv. Phys.},
  volume  = {56},
  pages   = {243--379},
  year    = {2007},
  doi     = {10.1080/00018730701223200}
}

@article{courteille1998observation,
  author  = {Courteille, Ph. and Freeland, R. S. and Heinzen, D. J. and van Abeelen, F. A. and Verhaar, B. J.},
  title   = {Observation of a Feshbach resonance in cold atom scattering},
  journal = {Phys. Rev. Lett.},
  volume  = {81},
  pages   = {69},
  year    = {1998},
  doi     = {10.1103/PhysRevLett.81.69}
}

@article{jaksch1998cold,
  author  = {Jaksch, D. and Bruder, C. and Cirac, J. I. and Gardiner, C. W. and Zoller, P.},
  title   = {Cold bosonic atoms in optical lattices},
  journal = {Phys. Rev. Lett.},
  volume  = {81},
  pages   = {3108},
  year    = {1998},
  doi     = {10.1103/PhysRevLett.81.3108}
}

@article{cazalilla2014ultracold,
  author  = {Cazalilla, M. A. and Rey, A. M.},
  title   = {Ultracold Fermi gases with emergent SU(N) symmetry},
  journal = {Rep. Prog. Phys.},
  volume  = {77},
  pages   = {124401},
  year    = {2014},
  doi     = {10.1088/0034-4885/77/12/124401}
}

@article{schreiber2015observation,
  author  = {Schreiber, M. and Hodgman, S. S. and Bordia, P. and L\"{u}schen, H. P. and Fischer, M. H. and Vosk, R. and Altman, E. and Schneider, U. and Bloch, I.},
  title   = {Observation of many-body localization of interacting fermions in a quasirandom optical lattice},
  journal = {Science},
  volume  = {349},
  pages   = {842--845},
  year    = {2015},
  doi     = {10.1126/science.aab3763}
}

@article{PhysRevResearch.5.L022010,
  author  = {Kiely, A. and O'Connor, E. and Fogarty, T. and Landi, G. T. and Campbell, S.},
  title   = {Entropy of the quantum work distribution},
  journal = {Phys. Rev. Res.},
  volume  = {5},
  pages   = {L022010},
  year    = {2023},
  doi     = {10.1103/PhysRevResearch.5.L022010}
}

@article{ref7,
  author  = {Aubry, S. and Andr\'{e}, G.},
  title   = {Analyticity breaking and Anderson localization in incommensurate lattices},
  journal = {Ann. Isr. Phys. Soc.},
  volume  = {3},
  pages   = {133},
  year    = {1980}
}

@article{ref8,
  author  = {Harper, P. G.},
  title   = {Single band motion of conduction electrons in a uniform magnetic field},
  journal = {Proc. Phys. Soc. London Sect. A},
  volume  = {68},
  pages   = {874},
  year    = {1955},
  doi     = {10.1088/0370-1298/68/10/304}
}

@article{PhysRevLett.123.180601,
  author  = {Mac\`{e}, N. and Alet, F. and Laflorencie, N.},
  title   = {Multifractal Scalings Across the Many-Body Localization Transition},
  journal = {Phys. Rev. Lett.},
  volume  = {123},
  pages   = {180601},
  year    = {2019},
  doi     = {10.1103/PhysRevLett.123.180601}
}

@article{PhysRevLett.119.260401,
  author  = {L\"{u}schen, H. P. and Bordia, P. and Scherg, S. and Alet, F. and Altman, E. and Schneider, U. and Bloch, I.},
  title   = {Observation of Slow Dynamics near the Many-Body Localization Transition in One-Dimensional Quasiperiodic Systems},
  journal = {Phys. Rev. Lett.},
  volume  = {119},
  pages   = {260401},
  year    = {2017},
  doi     = {10.1103/PhysRevLett.119.260401}
}

@article{PhysRevLett.115.180401,
  author  = {Mastropietro, V.},
  title   = {Localization of Interacting Fermions in the Aubry-Andr\'{e} Model},
  journal = {Phys. Rev. Lett.},
  volume  = {115},
  pages   = {180401},
  year    = {2015},
  doi     = {10.1103/PhysRevLett.115.180401}
}

@article{PhysRevA.95.033605,
  author  = {Settino, J. and Lo Gullo, N. and Sindona, A. and Goold, J. and Plastina, F.},
  title   = {Signatures of the single-particle mobility edge in the ground-state properties of Tonks-Girardeau and noninteracting Fermi gases in a bichromatic potential},
  journal = {Phys. Rev. A},
  volume  = {95},
  pages   = {033605},
  year    = {2017},
  doi     = {10.1103/PhysRevA.95.033605}
}

@article{PhysRevB.101.144303,
  author  = {Settino, J. and Talarico, N. W. and Cosco, F. and Plastina, F. and Maniscalco, S. and Lo Gullo, N.},
  title   = {Emergence of anomalous dynamics from the underlying singular continuous spectrum in interacting many-body systems},
  journal = {Phys. Rev. B},
  volume  = {101},
  pages   = {144303},
  year    = {2020},
  doi     = {10.1103/PhysRevB.101.144303}
}

@article{PhysRevB.87.134202,
  author  = {Iyer, S. and Oganesyan, V. and Refael, G. and Huse, D. A.},
  title   = {Many-body localization in a quasiperiodic system},
  journal = {Phys. Rev. B},
  volume  = {87},
  pages   = {134202},
  year    = {2013},
  doi     = {10.1103/PhysRevB.87.134202}
}

@article{bera2017one,
  author  = {Bera, S. and Martynec, T. and Schomerus, H. and Heidrich-Meisner, F. and Bardarson, J. H.},
  title   = {One-particle density matrix characterization of many-body localization},
  journal = {Ann. Phys. (Berlin)},
  volume  = {529},
  pages   = {1600356},
  year    = {2017},
  doi     = {10.1002/andp.201600356}
}

@article{lev2017transport,
  author  = {Lev, Y. B. and Kennes, D. M. and Kl\"{o}ckner, C. and Reichman, D. R. and Karrasch, C.},
  title   = {Transport in quasiperiodic interacting systems: From superdiffusion to subdiffusion},
  journal = {Europhys. Lett.},
  volume  = {119},
  pages   = {37003},
  year    = {2017},
  doi     = {10.1209/0295-5075/119/37003}
}

@article{naldesi2016detecting,
  author  = {Naldesi, P. and Ercolessi, E. and Roscilde, T.},
  title   = {Detecting a many-body mobility edge with quantum quenches},
  journal = {SciPost Phys.},
  volume  = {1},
  pages   = {010},
  year    = {2016},
  doi     = {10.21468/SciPostPhys.1.1.010}
}

@article{qxmt-m3rk,
  title = {Quantum coherence and anomalous work extraction in qubit gate dynamics},
  author = {Perciavalle, F. and Lo Gullo, N. and Plastina, F.},
  journal = {Phys. Rev. E},
  volume = {113},
  issue = {5},
  pages = {054126},
  numpages = {20},
  year = {2026},
  month = {May},
  publisher = {American Physical Society},
  doi = {10.1103/qxmt-m3rk},
  url = {https://link.aps.org/doi/10.1103/qxmt-m3rk}
}

@article{dl7m-8qyt,
  title = {Quantum coherence of topologically frustrated spin chains},
  author = {Ko\ifmmode \check{z}\else \v{z}\fi{}i\ifmmode \acute{c}\else \'{c}\fi{}, S. B. and Torre, G. and Deli\ifmmode \acute{c}\else \'{c}\fi{}, K. and Franchini, F. and Giampaolo, S. M.},
  journal = {Phys. Rev. B},
  volume = {112},
  issue = {14},
  pages = {144426},
  numpages = {11},
  year = {2025},
  month = {Oct},
  publisher = {American Physical Society},
  doi = {10.1103/dl7m-8qyt},
  url = {https://link.aps.org/doi/10.1103/dl7m-8qyt}
}

@article{Muniz2020,
  author  = {Muniz, Juan A. and Barberena, Diego and Lewis-Swan, Robert J. and Young, Dylan J. and Cline, Julia R. K. and Rey, Ana Maria and Thompson, James K.},
  title   = {Exploring dynamical phase transitions with cold atoms in an optical cavity},
  journal = {Nature},
  year    = {2020},
  volume  = {580},
  number  = {7805},
  pages   = {602--607},
  doi     = {10.1038/s41586-020-2224-x},
  url     = {https://doi.org/10.1038/s41586-020-2224-x},
  issn    = {1476-4687}
}

@article{Roati2008,
  author  = {Roati, Giacomo and D'Errico, Chiara and Fallani, Leonardo and Fattori, Marco and Fort, Chiara and Zaccanti, Matteo and Modugno, Giovanni and Modugno, Michele and Inguscio, Massimo},
  title   = {Anderson localization of a non-interacting Bose--Einstein condensate},
  journal = {Nature},
  year    = {2008},
  volume  = {453},
  number  = {7197},
  pages   = {895--898},
  doi     = {10.1038/nature07071},
  url     = {https://doi.org/10.1038/nature07071},
  issn    = {1476-4687}
}

@article{Greiner2002,
  author  = {Greiner, Markus and Mandel, Olaf and Esslinger, Tilman and H{\"a}nsch, Theodor W. and Bloch, Immanuel},
  title   = {Quantum phase transition from a superfluid to a Mott insulator in a gas of ultracold atoms},
  journal = {Nature},
  year    = {2002},
  volume  = {415},
  number  = {6867},
  pages   = {39--44},
  doi     = {10.1038/415039a},
  url     = {https://doi.org/10.1038/415039a},
  issn    = {1476-4687}
}

@article{Song2022,
  author  = {Song, Bo and Dutta, Shovan and Bhave, Shaurya and Yu, Jr-Chiun and Carter, Edward and Cooper, Nigel and Schneider, Ulrich},
  title   = {Realizing discontinuous quantum phase transitions in a strongly correlated driven optical lattice},
  journal = {Nature Physics},
  year    = {2022},
  volume  = {18},
  number  = {3},
  pages   = {259--264},
  doi     = {10.1038/s41567-021-01476-w},
  url     = {https://doi.org/10.1038/s41567-021-01476-w},
  issn    = {1745-2481}
}

@article{PhysRevLett.109.160601,
  title = {Emergent Thermodynamics in a Quenched Quantum Many-Body System},
  author = {Dorner, R. and Goold, J. and Cormick, C. and Paternostro, M. and Vedral, V.},
  journal = {Phys. Rev. Lett.},
  volume = {109},
  issue = {16},
  pages = {160601},
  numpages = {5},
  year = {2012},
  month = {Oct},
  publisher = {American Physical Society},
  doi = {10.1103/PhysRevLett.109.160601},
  url = {https://link.aps.org/doi/10.1103/PhysRevLett.109.160601}
}

@article{MatthiasVojta_2003,
doi = {10.1088/0034-4885/66/12/R01},
url = {https://doi.org/10.1088/0034-4885/66/12/R01},
year = {2003},
month = {nov},
publisher = {},
volume = {66},
number = {12},
pages = {2069},
author = {Matthias Vojta},
title = {Quantum phase transitions},
journal = {Reports on Progress in Physics}
}

@article{Palamara_2026,
doi = {10.1088/2058-9565/ae6a1b},
url = {https://doi.org/10.1088/2058-9565/ae6a1b},
year = {2026},
month = {may},
publisher = {IOP Publishing},
volume = {11},
number = {2},
pages = {025055},
author = {Palamara, Antonio and Plastina, Francesco and Sindona, Antonello and D’Amico, Irene},
title = {Full quantum work statistics for non-homogeneous many-body systems},
journal = {Quantum Science and Technology}
}

@article{PhysRevA.110.062203,
  title = {Thermal-density-functional-theory approach to quantum thermodynamics},
  author = {Palamara, Antonio and Plastina, Francesco and Sindona, Antonello and D'Amico, Irene},
  journal = {Phys. Rev. A},
  volume = {110},
  issue = {6},
  pages = {062203},
  numpages = {16},
  year = {2024},
  month = {Dec},
  publisher = {American Physical Society},
  doi = {10.1103/PhysRevA.110.062203},
  url = {https://link.aps.org/doi/10.1103/PhysRevA.110.062203}
}

@article{PhysRevB.89.165117,
  title = {Finite-temperature phase transitions in the ionic Hubbard model},
  author = {Kim, Aaram J. and Choi, M. Y. and Jeon, Gun Sang},
  journal = {Phys. Rev. B},
  volume = {89},
  issue = {16},
  pages = {165117},
  numpages = {8},
  year = {2014},
  month = {Apr},
  publisher = {American Physical Society},
  doi = {10.1103/PhysRevB.89.165117},
  url = {https://link.aps.org/doi/10.1103/PhysRevB.89.165117}
}

@article{AAA1,
author = {Zawadzki, Krissia and Canella, Guilherme A. and França, Vivian V. and D'Amico, Irene},
title = {Work Statistics and Entanglement Across the Fermionic Superfluid-Insulator Transition},
journal = {Advanced Quantum Technologies},
volume = {7},
number = {3},
pages = {2300237},
doi = {https://doi.org/10.1002/qute.202300237},
url = {https://advanced.onlinelibrary.wiley.com/doi/abs/10.1002/qute.202300237},
eprint = {https://advanced.onlinelibrary.wiley.com/doi/pdf/10.1002/qute.202300237},
year = {2024}
}

@article{PhysRevLett.105.095702,
  title = {Quantum Correlations in Spin Chains at Finite Temperatures and Quantum Phase Transitions},
  author = {Werlang, T. and Trippe, C. and Ribeiro, G. A. P. and Rigolin, Gustavo},
  journal = {Phys. Rev. Lett.},
  volume = {105},
  issue = {9},
  pages = {095702},
  numpages = {4},
  year = {2010},
  month = {Aug},
  publisher = {American Physical Society},
  doi = {10.1103/PhysRevLett.105.095702},
  url = {https://link.aps.org/doi/10.1103/PhysRevLett.105.095702}
}

@article{PhysRevA.111.012404,
  title = {Finite-temperature detection of quantum critical points via internal quantum teleportation},
  author = {Ribeiro, G. A. P. and Rigolin, Gustavo},
  journal = {Phys. Rev. A},
  volume = {111},
  issue = {1},
  pages = {012404},
  numpages = {16},
  year = {2025},
  month = {Jan},
  publisher = {American Physical Society},
  doi = {10.1103/PhysRevA.111.012404},
  url = {https://link.aps.org/doi/10.1103/PhysRevA.111.012404}
}

@article{PhysRevA.107.052420,
  title = {Detecting quantum critical points at finite temperature via quantum teleportation},
  author = {Ribeiro, G. A. P. and Rigolin, Gustavo},
  journal = {Phys. Rev. A},
  volume = {107},
  issue = {5},
  pages = {052420},
  numpages = {10},
  year = {2023},
  month = {May},
  publisher = {American Physical Society},
  doi = {10.1103/PhysRevA.107.052420},
  url = {https://link.aps.org/doi/10.1103/PhysRevA.107.052420}
}

@article{PhysRevA.83.062334,
  title = {Spotlighting quantum critical points via quantum correlations at finite temperatures},
  author = {Werlang, T. and Ribeiro, G. A. P. and Rigolin, Gustavo},
  journal = {Phys. Rev. A},
  volume = {83},
  issue = {6},
  pages = {062334},
  numpages = {10},
  year = {2011},
  month = {Jun},
  publisher = {American Physical Society},
  doi = {10.1103/PhysRevA.83.062334},
  url = {https://link.aps.org/doi/10.1103/PhysRevA.83.062334}
}

@article{doi:10.1142/S021797921345032X,
author = {WERLANG, T. and RIBEIRO, G. A. P. and RIGOLIN, GUSTAVO},
title = {INTERPLAY BETWEEN QUANTUM PHASE TRANSITIONS AND THE BEHAVIOR OF QUANTUM CORRELATIONS AT FINITE TEMPERATURES},
journal = {International Journal of Modern Physics B},
volume = {27},
number = {01n03},
pages = {1345032},
year = {2013},
doi = {10.1142/S021797921345032X},

URL = { 
    
        https://doi.org/10.1142/S021797921345032X
    
    

},
eprint = { 
    
        https://doi.org/10.1142/S021797921345032X
    
    

}
}

@article{Rodrigues2024,
  author  = {Rodrigues, Franklin L. S. and Lutz, Eric},
  title   = {Nonequilibrium thermodynamics of quantum coherence beyond linear response},
  journal = {Communications Physics},
  year    = {2024},
  volume  = {7},
  number  = {1},
  pages   = {61},
  doi     = {10.1038/s42005-024-01548-2},
  issn    = {2399-3650},
  date    = {2024-02-23}
}

@article{CAPPONI201650,
title = {Phases of one-dimensional SU(N) cold atomic Fermi gases—From molecular Luttinger liquids to topological phases},
journal = {Annals of Physics},
volume = {367},
pages = {50-95},
year = {2016},
issn = {0003-4916},
doi = {https://doi.org/10.1016/j.aop.2016.01.011},
url = {https://www.sciencedirect.com/science/article/pii/S0003491616000130},
author = {S. Capponi and P. Lecheminant and K. Totsuka}
}

@article{Domínguez-Castro_2019,
doi = {10.1088/1361-6404/ab1670},
url = {https://doi.org/10.1088/1361-6404/ab1670},
year = {2019},
month = {jun},
publisher = {IOP Publishing},
volume = {40},
number = {4},
pages = {045403},
author = {Domínguez-Castro, G A and Paredes, R},
title = {The Aubry–André model as a hobbyhorse for understanding the localization phenomenon},
journal = {European Journal of Physics}
}

@article{
doi:10.1126/science.aaa7432,
author = {Michael Schreiber  and Sean S. Hodgman  and Pranjal Bordia  and Henrik P. Lüschen  and Mark H. Fischer  and Ronen Vosk  and Ehud Altman  and Ulrich Schneider  and Immanuel Bloch },
title = {Observation of many-body localization of interacting fermions in a quasirandom optical lattice},
journal = {Science},
volume = {349},
number = {6250},
pages = {842-845},
year = {2015},
doi = {10.1126/science.aaa7432},
URL = {https://www.science.org/doi/abs/10.1126/science.aaa7432},
eprint = {https://www.science.org/doi/pdf/10.1126/science.aaa7432}}

\end{document}